# Are Graphene Sheets Stable?

N. García,
Laboratorio de Física de Sistemas Pequeños y Nanotecnología.
Consejo Superior de Investigaciones Científicas, Serrano 144,
Madrid -28006, Spain.

**The answer to the title question is yes and the sheets exhibit diffraction peaks but may not have long range crystalline order**. This is not a trivial question and answer and is immersing in the very active field now days of the study of properties of graphite and specially of graphene. This note is motivated in fact by a recent paper by Meyer et al (1) on the structure of suspended graphene sheets. This paper mentioned work by Landau(2) and Peierls(3) reporting that a 2D lattice melts and is instable due to thermal fluctuations and a later paper by Mermin(4) that prevents long range crystalline order for some atomic interactions. This is not easy to conclude and in fact for graphene its stability does not look so bad, in fact may have reasonable crystallinity.. The reasons are:

1) The works of Landau and Peierls have been subjected to further study that for example motivated the Mermin work. But this late work may not be applicable to graphene because of the stringent conditions needed in his proof that is valid for power-law potentials of the Lennard-Jones type. But not for potentials nonintegrable as the distance tends to zero and is inconclusive for core potentials. Where do we locate graphene?
2) Even in the case that the Mermin theorem is applicable to graphene the displacement autocorrelation function between two points in the sheet R and R´, variying as $\ln/R-R´/$, is very weak. And there is also directional long range order. Then this is not a liquid because the system should present local order within a given distance and still remain stable.
3) There is, at least, one paper concerning the problem by Imry and Gunther(5), not mentioned in Ref. 1, showing that even if the Debye-Waller factor exponent diverges and there is not strictly speaking long range crystalline order the system is stable, not a liquid, and what is more important the system exhibits local order in a given distance and diffraction peaks. This is due again to the fact that the divergences are logarithmic and in this territory the system is never a liquid below certain temperature. This is the same thing that for the isotropic 2D Heisenberg model, Mermin and Wagner proved (6) that the critical temperature is zero. But it turns out that tends to zero logarithmically with the anisotropy and then a very-very small anisotropy gives place to a reasonably large critical temperature as shown by theory and computer simulations (7).
4) Ref. 1 claims the existence of long range order in a suspended graphene sheet, I suppose that this is because diffraction peaks are observed or at least I have not seen in the paper other reasons, but this may not be due to long range crystalline order but to local crystalline order exhibiting diffraction peaks(5). Arguing that the sample behaves as a rough membrane neither proofs long range crystalline order because phenomena of local crystalline order of a given length ($\xi_T$) is also present in membranes (see the work of Nelson and Peliti(8)). However estimations give a $\xi_T \sim 10\mu m$ at RT, assuming the melting of grapheme to be ~1000K. This is a large region if one considers samples of the order of microns also.

Therefore the conclusions reached in Ref.1 of existence of long range order in their samples because behave as rough membranes are not so clear. *The existence of diffraction peaks are a necessary condition but are not sufficient.* My opinion from these problems is that, for very large samples, there are regions of good crystalline order separated by regions of disorder and meander paths where the strains are released. This could be resolved by diffraction experiments with beams of very large lateral coherence length that will correlate large areas of samples. This is important to know because the conduction properties of large graphene sheets may well became from interplay of these order-disorder regions. In graphite (HOPG) the conductivity depends of the conducting-less-conducting regions (9).

Finally I fully like the argument of Ref.1 that roughening and bending of the membrane will tend to make the system more crystalline. This has a physical explanation. All the arguments given above appear from infinite long wavelength phonons (energy tending to zero) that destroy long range crystalline order. However if we ripple the surface the phonons of wavelength larger than the ripple length will have a hard work to survive as such and the phonon spectrum should renormalize exhibiting a gap at infinite long wavelength. This gap recover crystallinity no matter how small is it as what happens with the anisotropy, producing a gap in the spin waves spectrum, in 2D Heisenberg model discussed above (7). Then, at the end the stability and ordering of graphene may look reasonable. For samples of the order of microns, there is another interesting point. To say, if the samples are of the order of microns then there is a cut-off in the phonons that help crystallinity, and also and more important, the order parameter estimated for local crystallinity is also of the order of microns. Then it will happen that most of the sample is crystalline and the strains creating disorder and other defects are located at the edges of the sample. Statistical sampling STM experiments on the sample looking at the edges in detail should resolve the problem.

I thank A.P. Levanyuk and P. Esquinazi for discussion and introducing me to the problem. This work has been supported by the EU-FP6 Ferrocarbon project.